\documentclass[]{spie}  

\usepackage{amsmath,amsfonts,amssymb}
\usepackage{graphicx}
\usepackage{physics}
\usepackage{multirow}
\usepackage{multicol}
\usepackage{booktabs}
\usepackage{adjustbox}
\usepackage[colorlinks=true, allcolors=blue]{hyperref}

\title{An Accelerated Pipeline for Multi-label Renal Pathology Image Segmentation at the Whole Slide Image Level}

\author[a]{Haoju Leng*}
\author[a]{Ruining Deng*}
\author[b]{Zuhayr Asad}
\author[c]{R. Michael Womick}
\author[d]{Haichun Yang}
\author[e]{Lipeng Wan}
\author[a]{Yuankai Huo}

\affil[a]{Department of Computer Science, Vanderbilt University, Nashville, TN, 37235 USA}
\affil[b]{College of Arts and Science, Vanderbilt University, Nashville, TN, 37235 USA}
\affil[c]{Department of Computer Science, The University of North Carolina at Chapel Hill, Chapel Hill, NC, 27514, USA}
\affil[d]{Department of Pathology, Vanderbilt University Medical Center, Nashville, TN 37215, USA}
\affil[e]{Department of Computer Science, Georgia State University, Atlanta, GA, 30302 USA}

\authorinfo{*Haoju Leng and Ruining Deng contribute equally to this paper\\  Corresponding author: Yuankai Huo: E-mail: yuankai.huo@vanderbilt.edu}

\usepackage{silence}
\begin{document} 
\maketitle

\begin{abstract}
Deep-learning techniques have been used widely to alleviate the labour-intensive and time-consuming manual annotation required for pixel-level tissue characterization. Our previous study introduced an efficient single dynamic network - Omni-Seg - that achieved multi-class multi-scale pathological segmentation with less computational complexity. However, the patch-wise segmentation paradigm still applies to Omni-Seg, and the pipeline is time-consuming when providing segmentation for Whole Slide Images (WSIs). In this paper, we propose an enhanced version of the Omni-Seg pipeline in order to reduce the repetitive computing processes and utilize a GPU to accelerate the model's prediction for both better model performance and faster speed. Our proposed method's innovative contribution is two-fold: (1) a Docker is released for an end-to-end slide-wise multi-tissue segmentation for WSIs; and (2) the pipeline is deployed on a GPU to accelerate the prediction, achieving better segmentation quality in less time. The proposed accelerated implementation reduced the average processing time (at the testing stage) on a standard needle biopsy WSI from 2.3 hours to 22 minutes, using 35 WSIs from the Kidney Tissue Atlas (KPMP) Datasets. The source code and the Docker have been made publicly available at \url{https://github.com/ddrrnn123/Omni-Seg}.


\end{abstract}


\section{INTRODUCTION}
\label{sec:intro}  


The Whole Slide Image (WSI) approach and other recent advanced techniques in digital pathology have led to a paradigm shift in the field. Digital pathology offers an opportunity to pathologists to work in front of remote monitors instead of local microscope~\cite{bengtsson2017computer}~\cite{marti2021digital}~\cite{gomes2021building}, and also sheds light on the possibility of computer-assisted quantification~\cite{madabhushi2016image}~\cite{eriksen2017computer}~\cite{kothari2013pathology}. 
Many pathological image segmentation approaches to pixel-level tissue characterization with deep-learning techniques have been proposed to alleviate labor-intensive and time-consuming manual annotation~\cite{lee2017pixel}. However, the heterogeneity of optimal scales for multi-tissue segmentation on renal histopathological images leads to the challenging issue of comprehensive semantic (multi-label) segmentation~\cite{graham2019hover}~\cite{srinidhi2021deep}~\cite{dabass2018review}~\cite{mahmood2019deep}. The prior arts are typically needed to train multiple segmentation networks to match the optimal pixel-resolution for heterogeneous tissue types (e.g., 5$\times$ for glomeruli, 10$\times$ for tubules, and 40$\times$ for capillaries in human renal pathology)~\cite{jayapandian2021development}. Moreover, large-scale giga-pixel pathological images segmentation requires large computational resource, restricting the deep learning models into patch-wise paradigm~\cite{lin2018scannet}~\cite{aresta2019bach}~\cite{rachmadi2017deep}.



To handle this issue, several deep learning models have been proposed within both multi-network~\cite{wang2020breast} or multi-head designs~\cite{chen2019med3d}. Our previous study introduced an efficient multi-label and multi-scale dynamic network - Omni-Seg~\cite{deng2022omni} - that achieved multi-class multi-scale pathological segmentation with less computational complexity. However, our proposed network still has multiple issues that can be ameliorated. The original Omni-Seg pipeline follows a patch-wise segmentation paradigm without the slide-wise merging process for the WSIs. This pipeline is also repetitive and time-consuming in regard to slide-wise segmentation due to the aggregation of tiled WSI patches at multiple scales. Using the intensity threshold to differentiate foreground and background can reduce the process time by differentiating the regions of tissue. However, setting a unified intensity threshold for WSIs is unreliable due to differences among digital scanning facilities and stain colors, thus failing to correctly localize the foreground tissues. Moreover, the prediction quality can still be improved because of the limitations of patch-wise segmentation and edge artifacts from Convolution Neural Networks (CNNs). There are still no efficient end-to-end tools for slide-wise multi-tissue multi-scale segmentation on WSIs in the pathology community.


In this paper, we propose an improved version of the Omni-Seg pipeline to reduce repetitive computing processes as well as utilize a GPU to accelerate the prediction for both better model performance and faster speed. Our proposed method's innovative contribution is two-fold: (1) a Docker is released for an end-to-end slide-wise multi-tissue segmentation for WSIs; and (2) the pipeline is deployed on a GPU to accelerate the prediction, achieving better segmentation quality in less time. The official implementation is available at \url{https://github.com/ddrrnn123/Omni-Seg}



\begin{figure}[t]
\begin{center}
\includegraphics[width=0.75\linewidth]{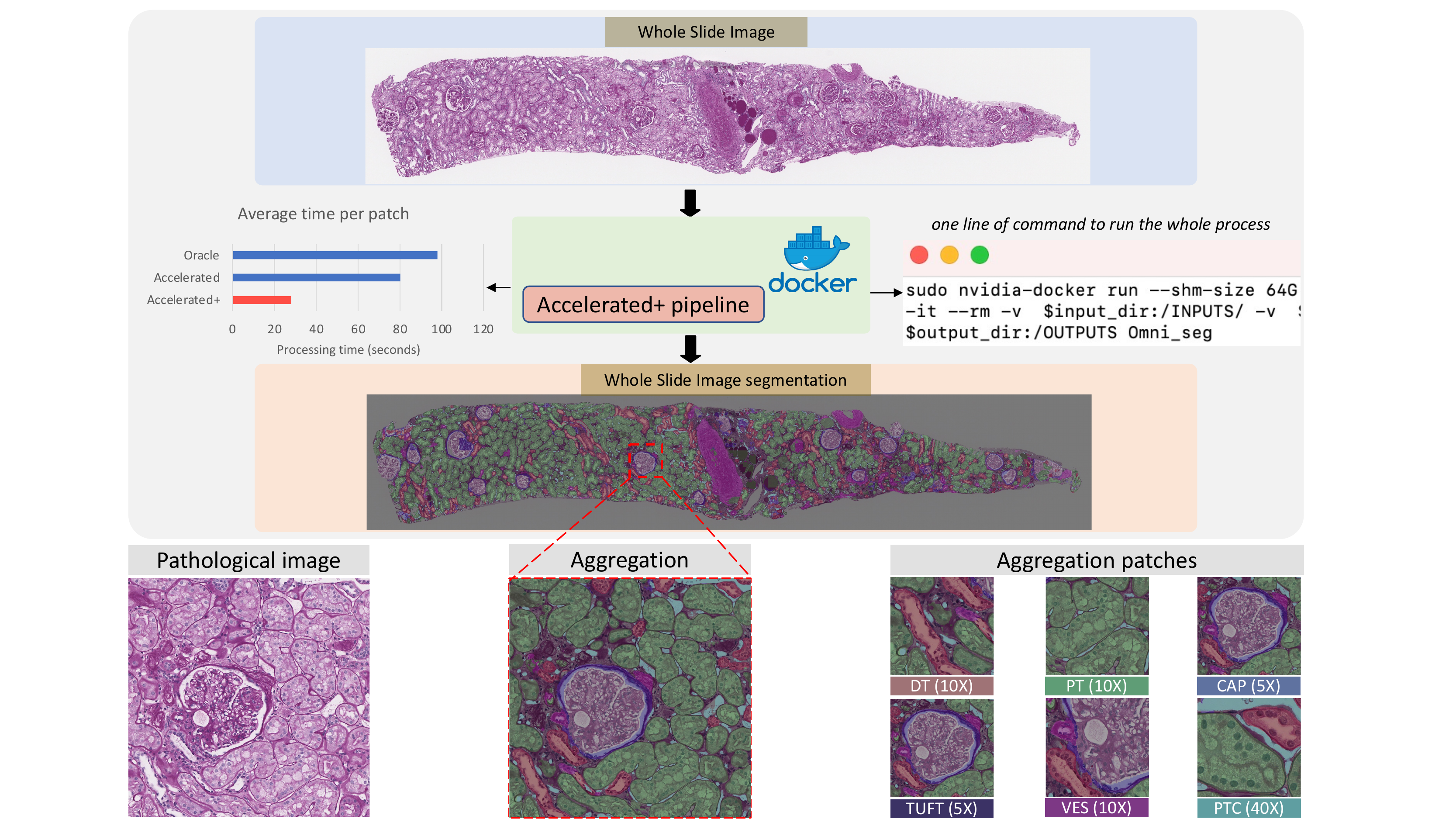}
\end{center}
   \caption{This figure presents an overview of the proposed pipeline. The input is the raw WSIs, while the output is the completely labeled tissue segmentation result. The domain impacts are to 1) enable slide-wise multi-tissue segmentation for WSIs via one command for non-technical users and to 2) achieve better segmentation quality with less time.}
\label{fig:problem}
 \end{figure}

\section{Method}

The proposed pipeline has three part: (1) WSI tissue detection and patch extraction; (2) Onmi-Seg segmentation; and (3) Slide-wise aggregation. 

\subsection{WSI tissue detection and patch extraction}

\begin{figure}[t]
\begin{center}
\includegraphics[width=0.65\linewidth]{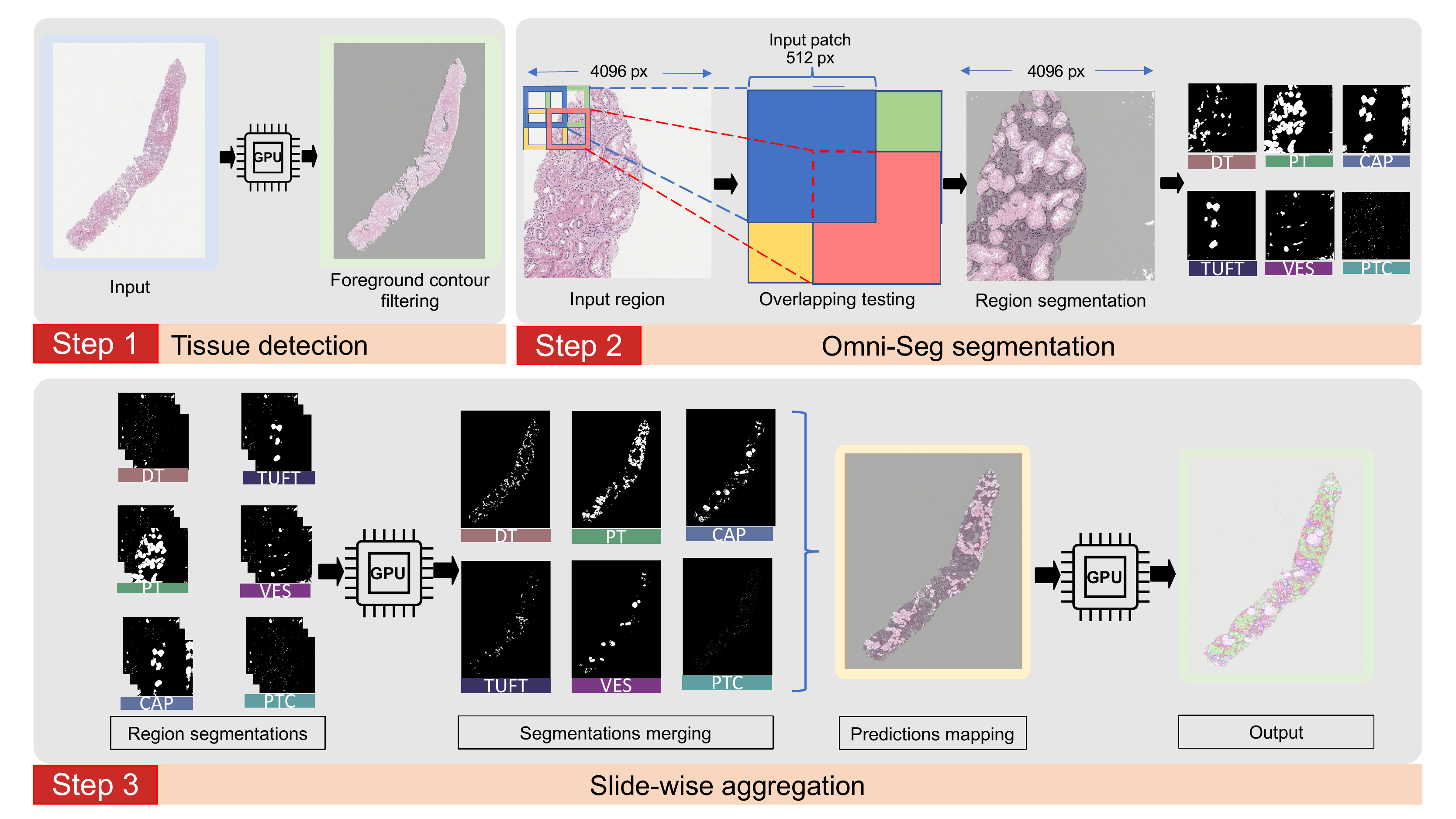}
\end{center}
   \caption{\textbf{Accelerated+ pipeline.} This figure shows the overall framework of the proposed pipeline. Step 1 represents tissue detection of the input. Step 2 represents the Omni-Seg segmentation of patches produced from Step 1. Step 3 represents the aggregation of the results from Step 2. Step 1 and 3 utilized a GPU to accelerate the process.}
   
\label{fig:pipeline}
 \end{figure}

The overall framework of the proposed pipeline is presented in Fig.\ref{fig:pipeline}. Padding was added to each WSI to handle the edge artifact issue. The original pipeline used 2048 pixels as the image padding size on its length and width, respectively, and the color of its padding is universal to different images.
However, WSIs could have backgrounds of significantly different colors due to the distinctive scanning machine settings and stain-color differences. In the proposed pipeline, we used padding size of 4096 pixels on its length and width. 
The color of the padding is determined by the mean of a $2\times2$ image vector on the top-left corner of the input image, which creates a particular padding color that is compatible with the background of the image. 

The size of the segmented image regions was $4096\times4096$ pixels. The regions were filtered by the foreground contour using an edge detection algorithm to remove blank regions that did not contain tissues. The foreground contour was set to a certain threshold to remove tiny noises from the image. To reduce processing time, the proposed pipeline saved image regions as numpy array data format (.npy) instead of the PNG data format. The proposed pipeline additionally utilized a GPU to perform extraction and selection steps whereas the original pipeline used CPU.


\subsection{Omni-Seg segmentation} 
We applied the pre-trained Omni-Seg model to selected patches. Each patch was tiled into small regions at three different scales (40$\times$, 10$\times$, and 5$\times$). Different magnifications of the tissue regions were used for predicting heterogeneous tissue types at optimal resolutions. To test different magnification patches, we applied the pre-trained Omni-Seg model to a small region size of $512\times512$ until it covered the whole patch, whereas the original pipeline used a size of $256\times256$ for region size. 

To increase segmentation accuracy, we used an overlap tiling strategy to segment regions. We set a certain stride smaller than the region size to overlap testing for certain times. The majority of the testing results of the regions would determine the final testing result of the patch. 
The predictions of various tissue types of the patch were separately saved as .npy data format in the proposed pipeline, instead of saving as .png data format in the original pipeline.

\subsection{Slide-wise Aggregation}
For each tissue type, the predictions of Omni-Seg were separately merged back into an image the original size of the input. The foreground contour generated during tissue detection was utilized again to filter out noises from the prediction of Omni-Seg. The 40$\times$ merged prediction was saved as the numpy data format instead of being repetitively saved as 40$\times$ and 10$\times$ predictions in a PNG data format, as was the case in the original pipeline. Then, all of the merged results of the various tissue types were stained with different colors on WSIs, which lead to the final prediction of a 40$\times$ image. A final downsampled prediction of the 10$\times$ WSI was also produced by directly resizing the final prediction of the 40$\times$ image for convenience, whereas in the original pipeline, the final downsampled prediction was stained from the 10$\times$ merged prediction. The proposed pipeline utilized a GPU – as opposed to a CPU used in the original pipeline – for all calculations during the merging and mapping processes in order to expedite the process.

\section{Data and Experimental design}
\subsection{Data}
Image patches were extracted from 35 kidney WSIs from the Kidney Tissue Atlas (KPMP) Datasets. 
The regions of interests - including glomerular tuft (TUFT), glomerular unit (CAP), proximal tubular (PT), distal tubular (DT), peritubular capillaries (PTC), and arteries (VES) - were manually annotated by the ImageScope software. All images were extracted from the 40$\times$ magnification of a WSI with the original pixel resolution of 0.25 Micron, and were saved in 40$\times$, 10$\times$ and 5$\times$ magnifications in the PNG format. Slides were stained with Hematoxylin and Eosin (H\&E), Periodic-acid-Schiff (PAS), Silver (SIL), Toluidine Blue (TOL), or Trichrome (TRI).


\subsection{Experimental Design}
For WSI tissue detection and patch extraction, the padding size was adjusted from 2048 pixels to 4096 pixels on length and width of the image respectively. The padding color was determined by mean of $2\times2$ image vector on top-left corner of WSI. The patch size was changed from 2048 pixels to 4096 pixels in the proposed pipeline. Ostu edge detection~\cite{xu2011characteristic} was used to receive foreground contour to remove any blank patch. Median blur was used on WSIs with threshold set to 59 to remove any tiny noise on the image for better contour detection. For the Omni-Seg testing stage, we splited the patch into $512\times512$ patches instead of $256\times256$ patches. The number of overlapping testing was reduced for different magnifications. 
For 40 $\times$ magnification testing, the stride was set to 256 pixels to reduce number of overlapping testing to 2 and save processing time, to have 2 tiles of overlapping testing and the major vote threshold was set to 1. For 10 and 5 $\times$ magnification testing, the stride was set to 64 pixels to have 8 tiles of overlapping testing, and the major vote threshold was set to 4. Due to reduced overlapping times and thus reduced computational complexity, 
We used GPU for the testing of all magnifications of the patch. For the merging and mapping process, the patch size was adjusted to 4096 pixels and stride size was adjusted to 2048 pixels. We utilized GPU for all calculations during the process. Instead of saving merged 40$\times$ and 10$\times$ prediction for different tissue types, the pipeline only saved merged 40$\times$ prediction for merging stage. Then, only the merged 40$\times$ prediction was mapped on the input image, instead of mapping 40$\times$ and 10$\times$ merged prediction separately. The 10$\times$ mapping result was from down-sampled 40$\times$ mapping result. We used NVIDIA RTX A5000 24G in the experiment.

\section{Results}
\begin{figure}
\begin{center}
\includegraphics[width=0.75\linewidth]{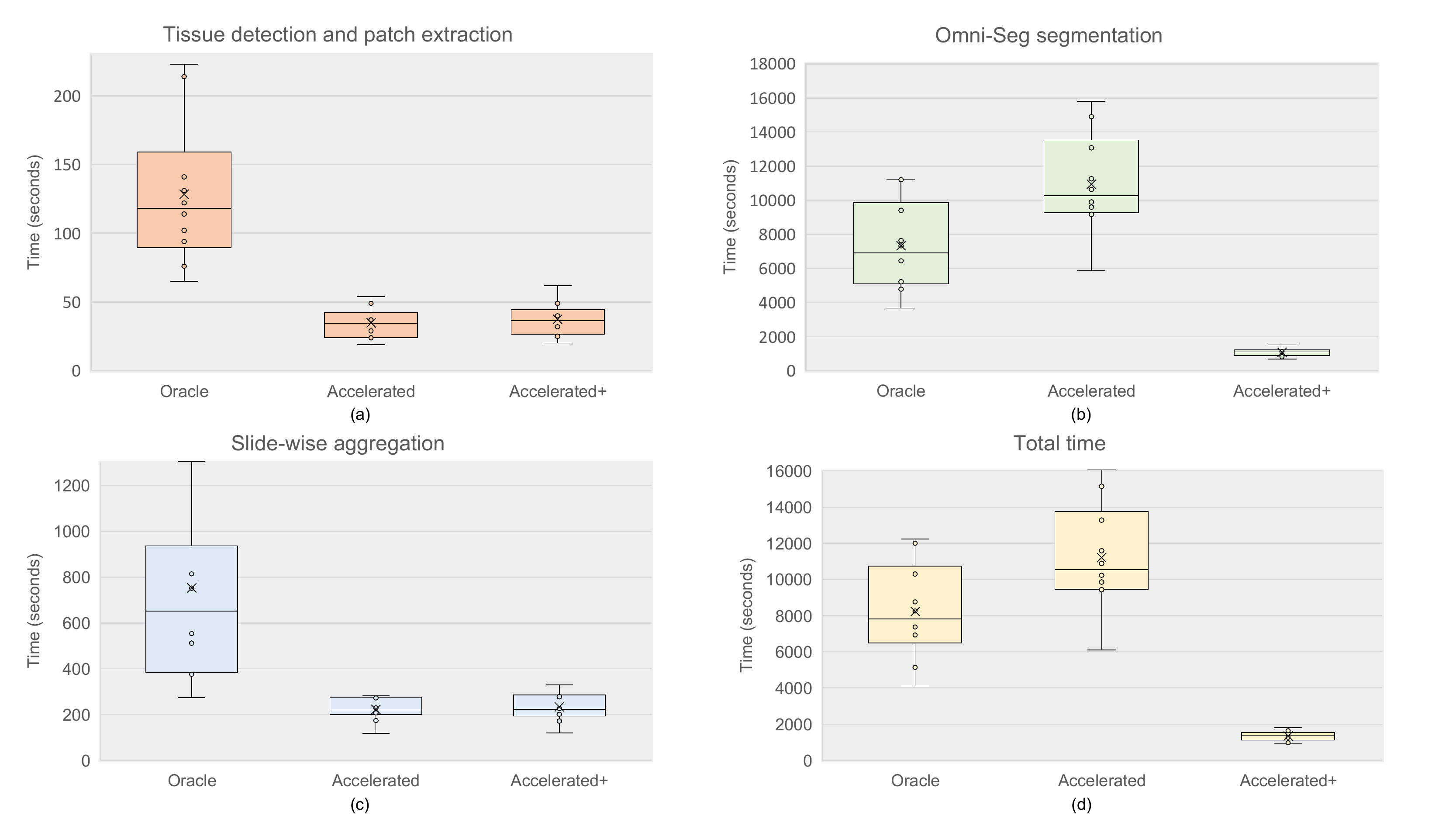}
\end{center}
   \caption{This figure shows the time needed for 10 WSIs in the Oracle, Accelerated, and Accelerated+ pipelines. (a) The time needed for the tissue detection and patch extraction process. (b) The time needed for Omni-Seg segmentation. (c) The time needed for the Slide-wise aggregation process. (d) The total time needed for the whole pipeline.}
\label{fig:speed}
 \end{figure}

Considering different versions of the Omni-Seg pipeline, we performed ten experiments using 10 kidney WSIs with the PAS stain on the Oracle, Accelerated, and the proposed Accelerated+ pipelines respectively. Fig.\ref{fig:speed} shows the completely labeled tissue segmentation result of the proposed pipeline. The processing time of the pipelines is demonstrated in the Table.\ref{table:performance}. The Table.\ref{table:performance} shows that the Accelerated+ pipeline uses less processing time on each patch. It also used less time for the Omni-Seg testing stage and patch segmentation merging and mapping, as compared to the Oracle and Accelerated pipelines. In addition, less image patches are cropped per each pathological image by the proposed Accelerated+ pipeline. 
Figure.\ref{fig:speed} also shows the processing time of the pipelines on 10 kidney WSIs with the PAS stain; as shown the figure, the Accelerated+ pipeline achieved the lowest median total processing time. 

\begin{table}[h]
\caption{Quantitative result of baseline methods.}
\centering
\begin{tabular}{l|ccc}
\toprule
 methods & Tissue detection+patch extraction(s)& Omni-Seg(s) & Slide-wise aggregation(s)\\
\midrule
 Oracle & 128.2 & 7341.6 & 449.9 \\
 Accelerated & \textbf{34.8} & 10948.8 & 27.8\\ 
 Accelerated+ & 37.4 & \textbf{1076.1} & \textbf{27.7}\\
\bottomrule
\end{tabular}

\begin{tabular}{l|ccc}
\toprule
 methods & Total time(s) & Total patches(s) & Time/patch(s)\\
\midrule
 Oracle & 82224 & 840 & 97.9\\
 Accelerated & 112146 & 1398 & 80.2\\ 
 Accelerated+ & \textbf{13213} & \textbf{478} & \textbf{28.2}\\
\bottomrule
\end{tabular}
\label{table:performance}
\end{table}

\section{Conclusion}
Per our results, the time for most of the modules and on each patch in our Accelerated+ pipeline significantly decreases with GPU usage. The total number of patches in the Oracle pipeline is less than that of the Accelerated pipeline, since the patch selection was not reliable since some patches were filtered out accordingly. The proposed method achieved less time on both the segmentation and aggregation processes, while subtly increasing the time for both tissue detection and extraction with a larger patch-size as compared to the Accelerated pipeline.


\section{ACKNOWLEDGMENTS}       
This work has not been submitted for publication or presentation elsewhere. 

\bibliography{main} 

\begin{thebibliography}{10}

\bibitem{bengtsson2017computer}
Bengtsson, E., Danielsen, H., Treanor, D., Gurcan, M.~N., MacAulay, C., and
  Moln{\'a}r, B., ``Computer-aided diagnostics in digital pathology,'' {\em
  Cytometry Part A}~{\bf 91}(6),  551--554 (2017).

\bibitem{marti2021digital}
Marti-Aguado, D., Rodr{\'\i}guez-Ortega, A., Mestre-Alagarda, C., Bauza, M.,
  Valero-P{\'e}rez, E., Alfaro-Cervello, C., Benlloch, S., P{\'e}rez-Rojas, J.,
  Ferr{\'a}ndez, A., Alemany-Monraval, P., et~al., ``Digital pathology:
  accurate technique for quantitative assessment of histological features in
  metabolic-associated fatty liver disease,'' {\em Alimentary Pharmacology \&
  Therapeutics}~{\bf 53}(1),  160--171 (2021).

\bibitem{gomes2021building}
Gomes, J., Kong, J., Kurc, T., Melo, A.~C., Ferreira, R., Saltz, J.~H., and
  Teodoro, G., ``Building robust pathology image analyses with uncertainty
  quantification,'' {\em Computer Methods and Programs in Biomedicine}~{\bf
  208},  106291 (2021).

\bibitem{madabhushi2016image}
Madabhushi, A. and Lee, G., ``Image analysis and machine learning in digital
  pathology: Challenges and opportunities,'' {\em Medical image analysis}~{\bf
  33},  170--175 (2016).

\bibitem{eriksen2017computer}
Eriksen, A.~C., Andersen, J.~B., Kristensson, M., dePont Christensen, R.,
  Hansen, T.~F., Kj{\ae}r-Frifeldt, S., and S{\o}rensen, F.~B.,
  ``Computer-assisted stereology and automated image analysis for
  quantification of tumor infiltrating lymphocytes in colon cancer,'' {\em
  Diagnostic pathology}~{\bf 12}(1),  1--14 (2017).

\bibitem{kothari2013pathology}
Kothari, S., Phan, J.~H., Stokes, T.~H., and Wang, M.~D., ``Pathology imaging
  informatics for quantitative analysis of whole-slide images,'' {\em Journal
  of the American Medical Informatics Association}~{\bf 20}(6),  1099--1108
  (2013).

\bibitem{lee2017pixel}
Lee, H., Troschel, F.~M., Tajmir, S., Fuchs, G., Mario, J., Fintelmann, F.~J.,
  and Do, S., ``Pixel-level deep segmentation: artificial intelligence
  quantifies muscle on computed tomography for body morphometric analysis,''
  {\em Journal of digital imaging}~{\bf 30}(4),  487--498 (2017).

\bibitem{graham2019hover}
Graham, S., Vu, Q.~D., Raza, S. E.~A., Azam, A., Tsang, Y.~W., Kwak, J.~T., and
  Rajpoot, N., ``Hover-net: Simultaneous segmentation and classification of
  nuclei in multi-tissue histology images,'' {\em Medical Image Analysis}~{\bf
  58},  101563 (2019).

\bibitem{srinidhi2021deep}
Srinidhi, C.~L., Ciga, O., and Martel, A.~L., ``Deep neural network models for
  computational histopathology: A survey,'' {\em Medical Image Analysis}~{\bf
  67},  101813 (2021).

\bibitem{dabass2018review}
Dabass, M., Vig, R., and Vashisth, S., ``Review of histopathological image
  segmentation via current deep learning approaches,'' in [{\em 2018 4th
  International Conference on Computing Communication and Automation
  (ICCCA)}{\nolinebreak\hspace{0.1em}]},   1--6, IEEE (2018).

\bibitem{mahmood2019deep}
Mahmood, F., Borders, D., Chen, R.~J., McKay, G.~N., Salimian, K.~J., Baras,
  A., and Durr, N.~J., ``Deep adversarial training for multi-organ nuclei
  segmentation in histopathology images,'' {\em IEEE transactions on medical
  imaging}~{\bf 39}(11),  3257--3267 (2019).

\bibitem{jayapandian2021development}
Jayapandian, C.~P., Chen, Y., Janowczyk, A.~R., Palmer, M.~B., Cassol, C.~A.,
  Sekulic, M., Hodgin, J.~B., Zee, J., Hewitt, S.~M., O’Toole, J., et~al.,
  ``Development and evaluation of deep learning--based segmentation of
  histologic structures in the kidney cortex with multiple histologic stains,''
  {\em Kidney international}~{\bf 99}(1),  86--101 (2021).

\bibitem{lin2018scannet}
Lin, H., Chen, H., Dou, Q., Wang, L., Qin, J., and Heng, P.-A., ``Scannet: A
  fast and dense scanning framework for metastastic breast cancer detection
  from whole-slide image,'' in [{\em 2018 IEEE winter conference on
  applications of computer vision (WACV)}{\nolinebreak\hspace{0.1em}]},
  539--546, IEEE (2018).

\bibitem{aresta2019bach}
Aresta, G., Ara{\'u}jo, T., Kwok, S., Chennamsetty, S.~S., Safwan, M., Alex,
  V., Marami, B., Prastawa, M., Chan, M., Donovan, M., et~al., ``Bach: Grand
  challenge on breast cancer histology images,'' {\em Medical image
  analysis}~{\bf 56},  122--139 (2019).

\bibitem{rachmadi2017deep}
Rachmadi, M.~F., Vald{\'e}s-Hern{\'a}ndez, M. d.~C., Agan, M. L.~F., and
  Komura, T., ``Deep learning vs. conventional machine learning: pilot study of
  wmh segmentation in brain mri with absence or mild vascular pathology,'' {\em
  Journal of Imaging}~{\bf 3}(4),  66 (2017).

\bibitem{wang2020breast}
Wang, Y., Lei, B., Elazab, A., Tan, E.-L., Wang, W., Huang, F., Gong, X., and
  Wang, T., ``Breast cancer image classification via multi-network features and
  dual-network orthogonal low-rank learning,'' {\em IEEE Access}~{\bf 8},
  27779--27792 (2020).

\bibitem{chen2019med3d}
Chen, S., Ma, K., and Zheng, Y., ``Med3d: Transfer learning for 3d medical
  image analysis,'' (2019).

\bibitem{deng2022omni}
Deng, R., Liu, Q., Cui, C., Yao, T., Long, J., Asad, Z., Womick, R.~M., Zhu,
  Z., Fogo, A.~B., Zhao, S., et~al., ``Omni-seg+: A scale-aware dynamic network
  for pathological image segmentation,'' (2022).

\bibitem{xu2011characteristic}
Xu, X., Xu, S., Jin, L., and Song, E., ``Characteristic analysis of otsu
  threshold and its applications,'' {\em Pattern recognition letters}~{\bf
  32}(7),  956--961 (2011).

\end{thebibliography}
\bibliographystyle{spiebib} 

\end{document}